%\magnification1200
\font\BBig=cmr10 scaled\magstep3
\font\small=cmr7

%%%%%%%%%%%%%%%%%%%%%%%%%%%%%%%%%%%%%%%%%%%
%%%%%%%%%%%%%% the title %%%%%%%%%%%%%%%%%%
%%%%%%%%%%%%%%%%%%%%%%%%%%%%%%%%%%%%%%%%%%%
\def\title{
{\bf\BBig
\centerline{The Dirac equation}
\bigskip
\centerline{in}\bigskip
\centerline{Taub-NUT space}
\bigskip
}}%%%%% for the front page

\def\runningtitle{
SUSY of Taub-NUT$\cdots$
} %%%%% for the header

%%%%%%%%%%%%%%%%%%%%%%%%%%%%%%%%%%%%%%%%%%%
%%%%%%%%%%%%% the author(s) %%%%%%%%%%%%%%%
%%%%%%%%%%%%%%%%%%%%%%%%%%%%%%%%%%%%%%%%%%%
\def\authors{
\centerline{A. COMTET\foot{Division de Physique Th\'eorique, IPN, Universit\'e
Paris-Sud, F-91405 ORSAY, Cedex (France) and
LPTPE, Universit\'e de Paris VI. 
}
and P. A. HORV\'ATHY
\foot{Laboratoire de Math\'ematiques et Applications,  
Facult\'e des Sciences, Parc de
Grandmont, Universit\'e, F-37200 TOURS (France). 
e-mail: horvathy@univ-tours.fr}
}}\bigskip

\def\runningauthors{Comtet and Horv\'athy
} %%%%% for the header

%%%%%%%%%%%%%%%%%%%%%%%%%%%%%%%%%%%%%%%%%%%
%%%%%%%%%%%%%% the metrics %%%%%%%%%%%%%%%%
%%%%%%%%%%%%%%%%%%%%%%%%%%%%%%%%%%%%%%%%%%%

\voffset = 1cm %%%%% ver printing
\baselineskip = 14pt %%%%% line spacing

\headline ={
\ifnum\pageno=1\hfill
\else\ifodd\pageno\hfil\tenit\runningtitle\hfil\tenrm\folio
\else\tenrm\folio\hfil\tenit\runningauthors\hfil
\fi\fi
} %%%%% header

\nopagenumbers
\footline = {\hfil} %%%%% footer

%%%%%%%%%%%%%%%%%%%%%%%%%%%%%%%%%%%%%%%%%%%
%%%%%%%%%%% some definitions %%%%%%%%%%%%%%
%%%%%%%%%%%%%%%%%%%%%%%%%%%%%%%%%%%%%%%%%%%

\def\ccr{\cr\noalign{\medskip}}
\def\parag{\hfil\break} %%%%% paragraph
\def\IR{{\bf R}} %%%%% the Reals
\def\and{\qquad\hbox{and}\qquad}

\def\smallover#1/#2{\hbox{$\textstyle{#1\over#2}$}}
\def\2{{\smallover 1/2}}
\def\D{D\mkern-2mu\llap{{\raise+0.5pt\hbox{\big/}}}\mkern+2mu}
\def\un{{\hat{\bf r}}} %% unit vector
\def\kikezd{\parag\underbar}
\def\o{{\rm o}}

%%%%%%%%%%%%%%%%%%%%%%%%%%%%%%%%%%%%%%%%%%%
%%%%%%%%%%%%%%% numberings %%%%%%%%%%%%%%%%
%%%%%%%%%%%%%%%%%%%%%%%%%%%%%%%%%%%%%%%%%%%

\newcount\ch %%%%% ch(apters
\newcount\eq %%%%% eq(uations
\newcount\foo %%%%% foo(tnotes
\newcount\ref %%%%% ref(erences

\def\chapter#1{
\parag\eq = 1\advance\ch by 1{\bf\the\ch.\enskip#1}
}

\def\equation{
\leqno(\the\ch.\the\eq)\global\advance\eq by 1
}

\def\foot#1{
\footnote{($^{\the\foo}$)}{#1}\advance\foo by 1
} %%%%% foot(notes

\def\reference{
\parag [\number\ref]\ \advance\ref by 1
}

\ch = 0 %%%%% global init ch(apter
%%%%% eq is set to 1 by \chapter
\foo = 1 %%%%% global init foo(tnote
\ref = 1 %%%%% global init ref(erence

%%%%%%%%%%%%%%%%%%%%%%%%%%%%%%%%%%%%%%%%%%%
%%%%%%%%%%%%%%%% the text %%%%%%%%%%%%%%%%%
%%%%%%%%%%%%%%%%%%%%%%%%%%%%%%%%%%%%%%%%%%%
\title
\vskip 1cm
\authors
\vskip 1cm
\parag
{\bf Abstract}
Using chiral supersymmetry, we show that 
the massless Dirac equation in the 
Taub-NUT gravitational instanton field is exactly soluble 
and explain the arisal and the use of the dynamical 
(super) symmetry.

\vskip1cm
\noindent{\it Physics Letters {\bf B 349}, p. 49-56 (1995)}

%%%%%%%%%%%%%%
\chapter{Introduction}
%%%%%%%%%%%%%%

The Dirac equation $\D\psi=0$ 
in the Kaluza-Klein (KK) monopole field 
--- obtained by imbedding the Taub-NUT solution 
into five-dimensional gravity with trivial time dependence [1] ---
was studied in the mid eighties [2].
It was found in particular that there were 
no zero modes and therefore the Rubakov effect was absent.
It was later realized, that the motion of a spin $0$ particle in
the KK monopole field is exactly soluble, 
and that it
has a Kepler-type dynamical symmetry [3], [4].  
Recently [5], a Runge-Lenz vector was constructed also in the
spinning case. 
Even more recently van Holten found, by applying the geometric 
technique of Ref. [6],
that the Taub-NUT 
gravitational instanton has an $N=4$ supersymmetry [7].
In this Letter, we point out that the fermion problem is also
{\it exactly soluble}, 
rederive the above symmetries following an
entirely different approach, and use them to find the spectrum and the
S-matrix.
The clue is {\it chiral supersymmetry} (SUSY): the Dirac operator, 
$
\D= 
\pmatrix{&T^{\dagger}\cr T&\cr}
$ is supersymmetric in $4$
dimensions: Its square,
$$
\D^2 
=\pmatrix{T^{\dagger}T&\cr&TT^{\dagger}\cr}
\equiv 
\pmatrix{H_1&\cr&H_0\cr}
\equiv
H,
\equation
$$
is in fact the Hamiltonian of a supersymmetric quantum mechanical 
system, with \lq fermion operator' 
$
\gamma^5=\pmatrix{1&\cr&-1\cr}
$
and intertwining transformations
$$
U={1\over\sqrt{H_0}}\,T
\and
U^{-1}\equiv U^{\dagger}=T^{\dagger}\,{1\over\sqrt{H_0}}.
\equation
$$
Indeed, $U^{\dagger}H_0 U=H_1$. 
\goodbreak
If $\psi$ is an $H_0$-eigenfunction with
eigenvalue $E^2\,> 0$, then  
$
\Psi=T^{\dagger}\psi
$
is an $H_1$-eigenfunction with the same eigenvalue. 
$H_1$ has therefore the same spectrum
as $H_0$ {\it up to zero-energy ground states} which
arise as solutions of $T\Psi=0$. If $H_0\psi=E^2\psi$, then 
$
\D\pmatrix{U^{\dagger}\psi\cr\pm\psi\cr}=\pm E\, 
\pmatrix{U^{\dagger}\psi\cr\pm\psi\cr},
$ 
so the eigenfunctions of the Dirac operator 
can be found from those of $\D^2$.

Let us assume that one of the partner hamiltonians, say $H_0$, is simple
so that its spectrum and symmetries are known.
Then the corresponding quantities for
$H_1$ are readily obtained by SUSY.
In the same spirit, if some operator $A_0$ is conserved for 
$H_0$ i.e. $[H_0, A_0]=0$, then 
$$
A_1=U^{\dagger}A_0U
\equation
$$
is conserved for $H_1$, $[H_1, A_1]=0$, and the diagonal operator
$
A={\rm diag}(A_1,A_0)
$
is conserved for the Dirac operator, $[\D, A]=0$. 
(With a slight abuse of notations, 
$A_1$ will be identified with ${\rm diag}(A_1,0)$, etc.).
The bosonic symmetries of
$\D$ can hence also be obtained from those of $H_1$
and have plainly 
the same algebraic structure as those of $H_0$. 
They can be, in principle,
calculated using Eq. (1.3). Their explicit forms are, however,
rather complicated as well as useless, since all their properties follow 
from those of $A_0$. They will therefore be omitted.

The supersymmetries of $H$ are easily identified and contain
$Q_1=\D$. Those which commute with $Q_1$ form the supersymmetry
algebra of the Dirac operator.
This
is precisely what happens
for the Taub-NUT metric: due to self-duality,
one of the partner Hamiltonians 
reduces to the {\it spin} $0$ Hamiltonian, 
and can be hence exactly solved. 
The other partner hamiltonian is
complicated, but SUSY dispenses us from solving it. 

%%%%%%%%%%%%%
\chapter{The Dirac operator in the Taub-NUT background}
%%%%%%%%%%%%%

The Taub-NUT gravitational instanton is described by the 4-metric
$$
V\Big\{dr^2+r^2(d\theta^2+\sin^2\theta d\phi^2)\Big\}
+
{1\over V}\Big\{d\psi-4m\vec{A}\cdot d\vec{r}\Big\}^2,
\equation
$$
where 
$$
V=1+{4m\over r}
\and
\vec{A}\cdot d\vec{r}=\cos\theta d\phi.
\equation
$$
In the usual context, the Taub-NUT
parameter, $m$ is positive; $m<0$ arises, e.g. 
in the long-range scattering of self-dual $SU(2)$
monopoles [3], [4].
The curved-space gamma matrices, $\hat\gamma^\mu$,
and the spin connection, $\Gamma^\mu$,
are expressed as [8]
$$
{\hat\gamma}^j=
\pmatrix{0&-
{i\over\sqrt{V}}\,\sigma^j\cr{i\over\sqrt{V}}\,\sigma^j&0\cr},
\qquad
{\hat\gamma}^4=\pmatrix{0 &\sqrt{V}-
{4im\over\sqrt{V}}\,\vec{\sigma}\cdot\vec{A}
\cr\sqrt{V}+{4im\over\sqrt{V}}\,\vec{\sigma}\cdot\vec{A}&0\cr},
\equation
$$
and
$$
\vec{\Gamma}=\pmatrix{
{4m\over 2V^2}(\vec{\nabla} V\cdot\vec{\sigma})\vec{A}
+{1\over 2V}\vec{\nabla} V\times\vec{\sigma}
\,&0\ccr0&0\cr},
\qquad
\Gamma_4=\pmatrix{-{1\over 2V^2}\,\vec{\nabla} V\cdot\vec{\sigma}
&0\ccr0&0\cr}, 
\equation
$$
respectively.
The operator $\partial_{\psi}$ commutes with the Dirac operator
$
\D=\hat\gamma^\mu(\partial_\mu+i\Gamma_\mu),
$
since nothing depends
on $\psi$.
Therefore, we can restrict ourselves
to a fixed eigensector of $\partial_{\psi}$.
Requiring $\psi$ to be periodic with period $16\pi m$, 
the eigenvalues of $-4mi\partial_{\psi}$,
denoted by $q$ and identified with the electric charge, 
are half-integers.
For a fixed $q$ (assumed positive for definiteness),
the Dirac operator 
$\D$
becomes
$$
\D=\pmatrix{0 &T^{\dagger}\cr T&0\cr}= 
\pmatrix{
0 &-\displaystyle{i\over\sqrt{V}}\vec{\sigma}\cdot\vec{\pi}
+{q\over 4m}\sqrt{V}
\cr&\cr\displaystyle{i\over V}\vec{\sigma}\cdot
\vec{\pi}\sqrt{V}+{q\over 4m}\sqrt{V} &0\cr}, 
\equation
$$
where
$
\vec{\pi}=-i{\bf\vec{\nabla}}+q\vec{A}
$. 
It is easy to check
that $T$ and $T^{\dagger}$ are each other's adjoint with respect to the
Taub-NUT volume element $V d^4 x$, as they have to be.
Using the (anti)self-duality property 
$\vec{\nabla} V=-4m\vec{B}$, the square of $\D$ is found to be (1.1) with
$$
\left\{\eqalign{
H_1&=H_0+{1\over V}
\Big[{q\over r^2V}\vec{\sigma}\cdot\un
+{4m\over r^2V}\vec{\sigma}\cdot\vec{r}\times\vec{L}_0
+\displaystyle{12m^2\over r^4V^2}\Big],
\ccr
H_0&={1\over V}
\Big[\vec{\pi}^2+({q\over4m})^2V^2\Big],
\cr}\right.
\equation
$$
where $\vec{L}_0=\vec{r}\times\vec{\pi}+q\un$ 
is the orbital angular momentum
[conserved for $H_0$ but not for $H_1$].
The partner hamiltonians $H_1$ and $H_0$ differ hence in a complicated
expression, and it is not at all obvious that they  
have the same spectra.
SUSY implies however that this is nevertheless true.
  
The \lq lower' (i.e. $\gamma^5=-1$)
sector  is simple:
all spin dependence dropped out. 
$H_0$ is in fact the 
the Hamiltonian for a spin $0$ particle in the KK field [3], [4] 
(times the unit matrix). 
Decomposing into radial and transverse components,
$H_0\Psi_0=E^2\Psi_0$ 
leads to the non-relativistic Coulomb-type equation
$$
\Big[-{1\over r^2}{\ d\over dr}\big(r^2{\ d\over dr}\big)
+{\vec{L}_0^2\over r^2} 
+\big({q^2\over 2m}-4mE^2\big){1\over r} 
+\big(({q\over 4m})^2-E^2\big)\Big]\Psi_0=0.
\equation
$$
Inserting $\vec{L}_0^2=L(L+1)$, $L=q,\, q +1,\ldots$
and setting
$$
\Psi_0^+=u_+(r)\,Y_L^{\mu}(\theta,\phi)\pmatrix{1\cr0\cr}
\and
\Psi_0^-=u_-(r)\,Y_L^{\mu}(\theta,\phi)\pmatrix{0\cr1\cr}, 
\equation
$$ 
where the $Y_L^{\mu}$'s are the \lq Wu-Yang' monopole harmonics,
we find that
both radial functions $u_{\pm}$ become
$$
u(r)=r^Le^{ikr}F(i\lambda+L+1,2L+2,-2ikr),
\equation
$$
where $F$ is the confluent hypergeometric function and 
$$
k^2=E^2-(q/4m)^2Ê
\and
\lambda=-4m\big[(E^2/2)-(q/4m)^2\big]/k.
\equation
$$
By Eq. (2.9), the
wave function  vanishes at the origin. No hermiticity problem
arises therefore.
Bound
states only arise for $E^2>(q/4m)^2$ which requires $m<0$. 
Square integrability
requires $\lambda^2=-(p+L+1)^2=-n^2$, $p=0, 1,\ldots$, yielding the
spectrum
$$
E_n^2={1\over8m^2}\sqrt{n^2-q^2}\Big(n-\sqrt{n^2-q^2}\Big),
\qquad n=q+1,\, q+2,\ldots 
\equation
$$
Such a state is labelled by the quantum numbers
$$
\left\{\matrix{
&E\qquad&\hbox{\small energy}
\ccr
&L=q,\,q+1,\ldots\qquad&\hbox{\small total orbital angular momentum}
\ccr
&\mu=-L,\ldots,L\qquad&\hbox{\small third component of}\ \vec{L}_0
\ccr
&s=\pm\2
&\hbox{\small eigenvalue of $\sigma_3$}\cr}\right.
\equation
$$
The $n^{th}$ energy level is therefore $2(n^2-q^2)$-fold degenerate.
By supersymmetry, the positive spectrum of the superpartner $H_1$
is again (2.11) with the same multiplicities; the eigenstates carry 
the same labels (2.12).
Zero modes only arise in the \lq upper' sector. Setting
$$
\psi_1=\pmatrix{
{e^{i(q/4m)\psi}\over\sqrt{V}}\chi
\ccr
0\cr},
\equation
$$
where $\chi$ is a two-component Pauli spinor,
the equation $\D\psi_1=0$ reduces to 
$$
\big[\vec{\sigma}\cdot\vec{\pi}-i(\smallover{q}/{4m})V\big]\chi=0.
\equation
$$
Applying the {\it flat-space} adjoint operator, 
$
\vec{\sigma}\cdot\vec{\pi}+i({q}/{4m})V,
$
yields  
$$
\Big[\vec{\pi}^2+2q{\vec{\sigma}\cdot\un\over r^2}
+\big({q\over4m}\big)^2
\big(1+{4m\over r}\big)^2\Big]
\chi=0,
\equation
$$
which is precisely the zero-mode equation of the
D'Hoker-Vinet \lq dyon' [9].
For $m>0$, there are no zero modes. 
This is consistent with the known results [2].
For $m<0$, Eq. (2.15) admits $2q$ normalizable solutions, 
described, e.g. in Ref. [10].
They have the lowest possible angular momentum, $j=q-\2$, 
and are expressed as 
$$
\chi=r^{q-1}e^{-qr}
\left(
\sqrt{q+\2-\mu\over2q+1}\,
Y_q^{\mu-1/2}\pmatrix{1\cr0\cr}
-
\sqrt{q+\2+\mu\over2q+1}\,
Y_q^{\mu+1/2}\pmatrix{0\cr1\cr}
\right),
\equation
$$
$\mu=-(q-\2),\ldots,(q-\2)$.
Scattering states arise for $E^2>(q/m)^2$. For a scalar particle, 
an incoming/outgoing state is characterized by its momentum
$\vec{k}_{in}\equiv \vec{k}$ and $\vec{k}_{out}\equiv\vec{k}'$ 
respectively. The scalar $S$-matrix reads [3]
$$
S_0(\vec{k}'|\vec{k})=
\sum_{L\geq q}(2L+1)
{(L-i\lambda)!\over(L+i\lambda)!}
{\cal D}^L_{-q,q}\big({\sl R}^{-1}(\vec{k}'){\sl R}(\vec{k})\big),
\equation
$$
where $\lambda$ is as in Eq. (2.10),
${\cal D}^L_{m,n}$ is the rotation matrix and
${\sl R}(\vec{k})$ denotes the rotation 
which takes the $z$-axis to the $\vec{k}$-direction.
In the spinning case [11], for the $H_0$ dynamics,  
incoming/outgoing states get an extra label, namely the third 
component of the spin, $s$ and $s'$, respectively. 
Since 
$s=\vec{k}\cdot\vec{\sigma}/2$,  
the $S$-matrix of $H_0$ is simply
$$
S(\vec{k}',s'|\vec{k},s)=S_0(\vec{k}'|\vec{k})\,
{\cal D}^{1/2}_{s',s}\big({\sl R}^{-1}(\vec{k}')
{\sl R}(\vec{k})\big).
\equation
$$
Now if $\psi_0^{in/out}$ is any incoming/outgoing wave for the $H_0$,
then $\psi_1=U(\psi_0^{in/out})$ is one for $H_1$ with the same
labels. Therefore,
the $S$-matrix for $H_1$ is again Eq. (2.18), just like
for $H_0$.
\goodbreak
%%%%%%%%%%%
\chapter{Bosonic symmetries}
%%%%%%%%

$H_0$ is conveniently viewed as the Hamiltonian of a spin $\2$ particle
---
but one with gyromagnetic ratio $g=0$. Being uncoupled,
the spin vector, 
$$
\vec{\rm S}_0=\2\vec{\sigma},
\equation
$$
is conserved. SUSY carries this over to $H_1$: 
$$
\vec{\rm S}_1=U^{\dagger}\vec{\rm S}_0U
\equation
$$ 
is conserved conserved for $H_1$. (The explicit expression is
not illuminating). The diagonal vector 
$
\vec{\rm S}={\rm diag}(\vec{\rm S}_1,\vec{\rm S}_0)
$
is therefore conserved for $\D$. Note that
$
\vec{\rm S}_1
$
and
$\vec{\rm S}_0$
are separately conserved for
$H={\rm diag}(H_1,H_0)$.

The Taub-NUT metric is radially symmetric. Therefore, the 
orbital angular momentum, 
$\vec{L}_0$, arises naturally when $H_0$ is viewed as 
describing two, uncoupled, spinless particles. 
But the total angular momentum,
$$
\vec{J}=\vec{L}_0+\vec{\rm S}_0,
\equation
$$
is also natural, when $H_0$ is viewed as describing
a spin$\2$ particle.
$H_0$ admits therefore {\it two}
independent conserved \lq angular momentum' operators.
The operators $\vec{\rm S}_0$ and $\vec{L}_0$ plainly commute,
so we get an $\o(3)\oplus\o(3)$ symmetry.
Again by supersymmetry, the superpartner, 
$H_1$, admits the same symmetries, namely $\vec{\rm S}_1$ in Eq. (3.2)
and
$$
\vec{L}_1=U^{\dagger}\vec{L}_0U.
\equation
$$
Since the total angular momentum, $\vec{J}$, is invariant,
$U^{\dagger}\vec{J}U=\vec{J}$, 
the partner of $\vec{L}_0=\vec{J}-\vec{\rm S}_0$ is also written as 
$
\vec{L}_1=\vec{J}-\vec{\rm S}_1=\vec{L}_0+\vec{\rm S}_0-\vec{\rm S}_1
$.
The Dirac operator $\D$ admits therefore the 
bosonic $\o(3)\oplus\o(3)$ symmetry, generated by
$$
\vec{\rm S}={\rm diag}(\vec{\rm S}_1,\vec{\rm S}_0)
\and
\vec{L}={\rm diag}(\vec{L}_1,\vec{L}_0).
\equation
$$
These operators were found by van Holten using a quite different method.
For example, $\vec{L}$ is his \lq improved angular momentum'.

$H_0$ actually admits further symmetries [3], 
namely a Runge-Lenz vector,
$$
\vec{K}_0=
\2\big\{\vec{\pi}\times\vec{L}_0-\vec{L}_0\times\vec{\pi}\big\} 
-4m\un\big(H_0-(\smallover{q}/{4m})^2\big).
\equation
$$
The vector operators $\vec{L}_0$ and $\vec{K}_0$ generate an $\o(3,1)$
dynamical symmetry for scattered motions and $\o(4)$ for bound
motions, 
to which $\vec{\rm S}_0$ adds an extra $\o(3)$. 

\goodbreak
SUSY transports the Runge-Lenz vector to $H_1$: 
$$
\vec{K}_1=U^{\dagger}\vec{K}_0U,
\equation
$$
and then to $\D$ to yield $\vec{K}={\rm diag}(\vec{K}_1,\vec{K}_0)$.
(The explicit form of $\vec{K}_1$ --- presented in Ref. [5] --- 
is {\it very} complicated).
The dynamical symmetry can be used to derive the 
bound-state spectrum and the S-matrix: for fixed energy $H=E^2$,
the Casimirs of the
full system satisfy, by construction, the same relations as
those of $H_0$, namely
$$\left\{\eqalign{
&{\vec{K}\over\sqrt{E^2-(q/4m)^2}}\cdot\vec{L}=
-4mq{E^2/2-(q/4m)^2\over\sqrt{E^2-(q/4m)^2}},
\ccr
&\Big({\vec{K}\over\sqrt{E^2-(q/4m)^2}}\Big)^2
-\vec{L}^2=
1-q^2
+\Big(4m{E^2/2-(q/4m)^2\over\sqrt{E^2-(q/4m)^2}}\Big)^2,
\ccr
&\big(\vec{\rm S}\big)^2=\smallover3/4,
\cr}\right.
\equation
$$
confirming that the spin is indeed decoupled.
From the representation theory of $\o(4)$
we infer that
$$
4m{({q/4m})^2-E^2/2\over\sqrt{(q/4m)^2-E^2}}=n,
\qquad
n=q+1,\ldots
\equation
$$
yielding the spectrum (2.11) once again.
The scattering can be described using the spinning extension of
Zwanziger's approach [11]; one finds Eq. (2.18) once again. 
It is worth stressing that the only effect of spin is
to double the multiplicity for bound states, and to multiply the 
$S$ matrix by the spin factor, respectively.
\goodbreak
\chapter{Supersymmetry}[12]

Consider first the supersymmetries of $H=\D^2$, restricted to
positive-energy states.
Firstly, the two odd scalar charges
$$
Q_1=\pmatrix{0&T^\dagger\cr T&0\cr},
\qquad
Q_2=-i\gamma^5Q_1=\pmatrix{0&-iT^\dagger\cr iT&0\cr}
\equation
$$
are both square-roots of $H$,
$$
\{Q_\alpha,Q_\beta\}=2\delta_{\alpha\beta}H.
\equation
$$

Now the components of the two \lq spins', $\vec{\rm S}_0$ and 
$\vec{\rm S}_1$, span two independent $\o(3)\simeq{\rm su}(2)$'s,
$$
[{\rm S}_a^i,{\rm S}_b^j]=i\epsilon^{ij}_k\delta_{ab}\,
{\rm S}_a^k,
\equation
$$
($a,\,b=0,1$) as well as commute with $\gamma^5$,
$$
[\gamma^5,\vec{\rm S}_a]=0.
\equation
$$
But they also satisfy the anticommutation relation
$
\{{\rm S}_a^i,{\rm S}_b^j\}=(1/2)\,\delta^{ij}\delta_{ab};
$
therefore, the following two vector odd charges
$$
\vec{Q}_\alpha=2i\,[\vec{\rm S}_0,Q_\alpha]
\qquad(\alpha=1,2)
\equation
$$
also commute with $H=\D^2$. 
The commutation relations are (4.3)-(4.4), augmented with
$$
\matrix{
[\gamma^5,Q_\alpha]=2i\epsilon_{\alpha\beta}Q_\beta,\hfill
&[\gamma^5,Q_\alpha^k]=2i\epsilon_{\alpha\beta}Q_\beta^k,\qquad\hfill
\ccr
[S_0^i,Q_\alpha^j]=\2i\big(
\delta_{ij}Q_\alpha+\epsilon_{ijk}Q_\alpha^k\big),\qquad\hfill
&[S_1^i,Q_\alpha^j]=-\2i\big(
\delta_{ij}Q_\alpha-\epsilon_{ijk}Q_\alpha^k\big),\hfill
\ccr
[\vec{\rm S}_0,Q_\alpha]=-\2i\vec{Q}_\alpha,\hfill
&[\vec{\rm S}_1,Q_\alpha]=\2i\vec{Q}_\alpha.\hfill
\cr}
\equation
$$
The anticommutation relations read in turn (4.2) and
$$\eqalign{
&\{Q_\alpha,\vec{Q}_\beta\}=-4H\,\epsilon_{\alpha\beta}(\vec{\rm S}_1+\vec{\rm S}_0),
\cr
&\{Q_\alpha^i,Q_\beta^j\}=2H\,\delta_{\alpha\beta}\delta_{ij}
-4H\epsilon_{\alpha\beta}\epsilon_{ijk}(S_1^k-S_0^k).
\cr}
\equation
$$
Since all odd operators anticommute with $\gamma^5$,
we conclude that $H=\D^2$ admits the
superalgebra ${\rm u}(2/2)$ as symmetry. 
It is easy to read off the symmetries of
$\D\equiv Q_1$ from these relations: 
$Q_1$ itself and
$$
\vec{Q}_2=
\pmatrix{0&{i\over\sqrt{V}}\big(\vec{\pi}
+i\vec{\sigma}\times\vec{\pi}\big)
-\big({q\over4m}\big)\sqrt{V}\vec{\sigma}
\ccr
-{i\over V}\big(\vec{\pi}-i\vec{\sigma}\times\vec{\pi}\big)\sqrt{V}
-\big({q\over4m}\big)\sqrt{V}\vec{\sigma}
&0\cr}
\equation
$$
generate, with 
$\vec{\rm S}={\rm diag}(\vec{\rm S}_1,\vec{\rm S}_2)$, 
an $N=4$ sub-superalgebra:
$$
\matrix{
[S^i,Q_2^j]=i\epsilon_{ijk}Q_2^k,\qquad\hfill
&[\vec{\rm S},Q_1]=0,\hfill
&\hfill
\ccr
\{Q_1,Q_1\}=2H,\hfill
&\{Q_2^i,Q_2^j\}=2H\delta_{ij},\qquad\hfill
&\{Q_1,\vec{Q}_2\}=-4H\vec{\rm S}.\hfill
\cr}
\equation
$$
In Ref. [7], these supercharges are associated to Killing-Yano tensors.

\goodbreak
\chapter{Discussion}

Some of the results above can be generalized.
The multi-Taub-NUT metric is given, for example,
again by (2.1), but (2.2) replaced rather by
$$
V=1+\sum_i{4m\over|\vec{r}-\vec{r}_i|}.
\equation
$$
Requiring 
(anti) self-duality, 
$\vec{\nabla} V=-4m\vec{B}$, we still obtain a solution of the
vacuum Einstein equations.
The square of $\D$ is (1.1) with $H_0$ as in Eq. (2.6)
(while $H_1$ is a complicated expression similar to the one before).
The spin drops out again in the 
\lq lower' sector, giving rise to the same supersymmetry algebra
as above 
(while the $\o(4)/\o(3,1)$ Kepler symmetry is in general broken).
Similar statements should hold for other self-dual metrics like
the Eguchi-Hanson or Atiyah-Hitchin metric, etc.

Unlike in Ref. [5],
our conserved electric charge $q$ contains no spin contribution. 
The explanation may well be the following one.
Taub-NUT space has the geometry of $S^3\times\IR^+$;
the four-dimensional rotation group, 
${\rm SO}(4)$, acts on the $S^3$ part by isometries. 
The $\partial_\psi={\rm const}.$ condition breaks
${\rm SO}(4)$ to ${\rm SO}(3)\times{\rm SO}(2)$; 
the conserved quantities
associated to the ${\rm SO}(3)$ (respectively to ${\rm SO}(2)$) are the 
total angular momentum, $\vec{J}$,
and the electric charge, $q$, respectively. 
The conserved quantities can, however, be 
\lq improved' [7]. For the ${\rm SO}(3)$, this yields the \lq orbital'
expressions
$\vec{L}=\vec{J}-\vec{S}$. It is likely that the spin part of the
\lq charge' of Ref. [5] can be removed by a similar \lq improvement', 
leaving us with the spin-independent $-i\partial_\psi$.

In the \lq dyon' case, an additional conserved quantity is provided by
the analog of Dirac's operator [10],
$$
{\cal K}=
\pmatrix{
0&\vec{\sigma}\cdot\vec{\ell}+1\hfill
\cr
-(\vec{\sigma}\cdot\vec{\ell}+1)\hfill&0
\cr},
\qquad 
\vec{\ell}=\vec{r}\times\vec{\pi}.
\equation
$$
${\cal K}$
commutes with the Dirac operator %$[\D,{\cal K}]=0$. 
and is 
associated to a Penrose-Floyd tensor [13]. It has the remarkable 
property, that its anticommutator with the vector supercharge 
$\vec{Q}_2$
reproduces the Runge-Lenz vector. More precisely,
$$
{1\over2}\big\{\vec{Q}_2,{\cal K}\big\}
=\vec{K}
-q(\vec{J}+\vec{\rm S}).
\equation
$$

The results of van Holten indicate, that a similar statement
holds in the
Taub-NUT case also. We have, however, not yet found the curved-space
version of the ${\cal K}$ in Eq. (5.2); the obvious guess
$\sqrt{V}{\cal K}$ neither commutes with $\D$ nor agrees with Eq. (5.3)
completely.

Another intriguing problem is to extend the bosonic $\o(4,2)$ 
symmetry [3, 4] to the spinning case.

\kikezd{Acknowledgement}.
We are indebted to L. Feh\'er for enlightening discussions and
for allowing us to use his unpublished notes. 
We thank also J. W. van Holten for correspondence.

\vskip3mm
\goodbreak
%%%%%%%%%%%%%%%%%%%%%%%%%%%%%%%%%%%%%%%%%%%
%%%%%%%%%%%%% the references %%%%%%%%%%%%%%
%%%%%%%%%%%%%%%%%%%%%%%%%%%%%%%%%%%%%%%%%%%

\centerline{\bf References}

\reference
D. Gross and M. J. Perry, Nucl. Phys. {\bf B226}, 29 (1983);
R. Sorkin, Phys. Rev. Lett. {\bf 51}, 87 (1983).

\reference
Z. F. Ezawa and A. Iwazaki, Phys. Lett. {\bf 138B}, 81 (1984);
M. Kobayashi and A. Sugamoto, Prog. Theor. Phys. {\bf 72}, 122 (1984);
A. Bais and P. Batenburg Nucl. Phys. {\bf B245}, 469 (1984).

\reference
G. W. Gibbons and N. Manton,
Nucl. Phys. {\bf B274}, 183 (1986);
G. W. Gibbons and P. Ruback,
Phys. Lett. {\bf 188B}, 226 (1987);
Comm. Math. Phys. {\bf 115}, 267 (1988).

\reference
L. Feh\'er and P. A. Horv\'athy,
Phys. Lett. {\bf 183B}, 182 (1987); 
B. Cordani, L. Feh\'er and P. A. Horv\'athy,
Phys. Lett. {\bf 201B}, 481 (1988)
T. Iwai and Katayama, Journ. Geom. Phys. {\bf 12}, 55 (1993).

\reference
M. Visinescu,
Phys. Lett. {\bf B339}, 28 (1994);
Class. Quant. Grav. {\bf 11}, 1867 (1994).

\reference
G. W. Gibbons, R. H. Rietdijk and J. W. van Holten,
Nucl. Phys. {\bf B404}, 42 (1993).

\reference
J. W. van Holten,
Phys. Lett. {\bf B342}, 47 (1995).

\reference
H. Boutaleb Joutei,
A. Chakrabarti and A. Comtet,
Phys. Rev. {\bf D21}, 2280 (1980)

\reference
E. D'Hoker and L. Vinet, 
Phys. Rev. Lett. {\bf 55}, 1043 (1986).

\reference
F. Bloore and P. A. Horv\'athy,
Journ. Math. Phys. {\bf 33}, 1869 (1992);
M. Berrondo and H. V. McIntosh, Journ. Math. Phys. {\bf 11}, 125 (1970).

\reference
D. Zwanziger, Phys. Rev. {\bf 176}, 1480 (1968);
L. Gy. Feh\'er and P. A. Horv\'athy,
Mod. Phys. Lett {\bf A3}, 1451 (1988).

\reference
L. Gy. Feh\'er,
unpublished notes (1988);
L. Gy. Feh\'er, P. A. Horv\'athy and L. O'Raifeartaigh,
Int. Journ. Mod. Phys. {\bf A4}, 5277 (1989). 

\reference
S. Durand, J.-M. Lina and L. Vinet,
Lett. Math. Phys. {\bf 17}, 289 (1989).

\vfill\eject
\bye